\documentclass[a4paper,12pt]{article}
\usepackage{amssymb,amsmath,mathrsfs}
\usepackage[usenames,dvipsnames]{xcolor}
\usepackage{hyperref}
\usepackage{tensor}
\usepackage{graphicx}
 \usepackage[LGR,T1]{fontenc} 
 \usepackage[utf8]{inputenc}
\newcommand{\textgreek}[1]{\begingroup\fontencoding{LGR}\selectfont#1\endgroup}
\usepackage[left=1.2in,top=1in,right=1.2in,bottom=1in,headheight=0.8in,foot=0.5in]{geometry}
\usepackage[nodisplayskipstretch]{setspace} 
\usepackage[width=\textwidth]{caption}
\usepackage{pifont}% http://ctan.org/pkg/pifont

\newcommand{\nocontentsline}[3]{}
\newcommand{\tocless}[2]{\bgroup\let\addcontentsline=\nocontentsline#1{#2}\egroup}
\hypersetup{
        breaklinks=true,
	colorlinks=true,         
	linkcolor=Black,          
	citecolor=MidnightBlue,
	urlcolor=MidnightBlue            
 }
\usepackage[indentafter]{titlesec}
\titleformat{name=\section}{}{\thetitle.}{0.8em}{\centering\scshape}
\titleformat{name=\subsection}[runin]{}{\thetitle.}{0.5em}{\bfseries}[.]
\titleformat{name=\subsubsection}[runin]{}{\thetitle.}{0.5em}{\itshape}[.]
\titleformat{name=\paragraph,numberless}[runin]{}{}{0em}{}[.]
\titlespacing{\paragraph}{0em}{0em}{0.5em}
\titleformat{name=\subparagraph,numberless}[runin]{}{}{0em}{}[.]
\titlespacing{\subparagraph}{0em}{0em}{0.5em}

\usepackage[style=ext-authoryear-comp,maxcitenames=2,uniquename=false,backend=biber,uniquelist=false,articlein=false]{biblatex}
\DeclareNameAlias{sortname}{last-first}

\title{The Virtues of Pursuit-Worthy Speculation: The Promises of Cosmic Inflation\footnote{Forthcoming in \textit{British Journal for the Philosophy of Science}}}

\author{William J.~Wolf\footnote{Faculty of Philosophy, University of Oxford, UK. william.wolf@philosophy.ox.ac.uk}, \& Patrick M. Duerr\footnote{Martin Buber Society of Fellows for Research in the Humanities and Social Sciences, Hebrew University of Jerusalem, IL. Email:\  patrick-duerr@gmx.de.}\footnote{Faculty of Philosophy, University of Oxford, UK. Email: patrick.duerr@philosophy.ox.ac.uk.}}

\date{}
\makeatletter
\let\uppercasenonmath\@gobble

\begin{document}
\setstretch{1.0}
\maketitle
% \vspace{-.5in}

% \begin{center}
% \author{William J.~Wolf\footnote{Faculty of Philosophy, University of Oxford, UK. Email:\ william.wolf@philosophy.ox.ac.uk, williamjwolf2@gmail.com.} and Patrick M. Duerr\footnote{Martin Buber Society of Fellows for Research in the Humanities and Social Sciences, Hebrew University of Jerusalem, IL. Email:\  patrick-duerr@gmx.de.}\footnote{Faculty of Philosophy, University of Oxford, UK. Email: patrick.duerr@philosophy.ox.ac.uk.}}
% \end{center}\vspace{.1in}

\begin{abstract}
\noindent
The paper investigates the historical and contemporary pursuit-worthiness of cosmic inflation---the rationale for \textit{working on} it (rather than necessarily the evidential support for claims to its approximate \textit{truth}): what reasons existed, and exist, that warrant inflation’s status as the mainstream paradigm studied, explored, and further developed by the majority of the cosmology community? We’ll show that inflation exemplifies various salient theory virtues: explanatory depth, unifying/integrative power, fertility and positive heuristics, the promotion of understanding, and the prospect (and passing) of novel benchmark tests. This, we'll argue, constitutes inflation’s auspicious promise. It marks inflation as preferable over both the inflation-less Hot Big Bang Model, as well as rivals to inflation: inflation, we maintain, rightly deserved, and continues to deserve, the concerted research efforts it has enjoyed.


\end{abstract}
\tableofcontents
\setstretch{1.2}

\section{Introduction}

\noindent The theory of inflation, proposed in the early 1980s \parencite{Guth:1980zm, Starobinsky:1980te}, modifies the classical Big Bang model within the first tiny fraction of a second after the Big Bang: it  postulates an exotic type of matter, the ``inflaton”, that caused the fledgling universe to undergo a dramatic phase of exponential growth---``cosmic inflation”. This was supposed to remedy some of the perceived defects that beset the then-standard Hot Big Bang model (henceforth ``HBB-model''), based on General Relativity and ordinary\footnote{For the purposes of the present paper, ``ordinary matter” shouldn’t be contrasted with ``\textit{baryonic} matter” (i.e.\ matter composed of quarks), as is customary in another cosmological/astrophysical contexts---the Dark Matter problem (e.g.\ \textcite{Bertone:2016nfn}). What makes the matter involved in inflation exotic is that it violates certain so-called energy conditions; these are typically viewed as general, high-level constraints which matter, as it figures in classical field theory, should respect (see e.g.\ \textcite[Ch.5]{Malament2012-MALTIT}). }  matter (\textit{sans} the inflaton). The received model wasn't suffering from unimpeachable \textit{empirical} inadequacies; rather it was faulted with achieving its adequacy by dint of artificiality or contrivance: only through conspicuous choice of special initial conditions could certain observational peculiarities be accounted for. The advocates of inflation touted it as proffering ``a more convincing story'' \parencite[p.34]{guth1989inflationary}.

Not everyone has been swayed, however. The theory has discomfited many who have questioned its methodological standing (see \textcite{Wolf:2022yvd, DawidManuscript-DAWIIC} for recent philosophical discussions comparing inflation and alternative bouncing cosmologies). Although inspired by speculative ideas in high-energy physics and propounded while no empirical evidence in its support seemed forthcoming, inflation took the cosmology community by storm. Inflation was quickly elevated to a mainstream area of cosmological research; within a few years it became a key element in the current standard model of cosmology. Yet, observations in its favour would have to wait for roughly another 20 years. Even that assessment requires a non-trivial pinch of sanguinity: still today it remains contentious \textit{how compelling} the empirical case for inflation is.

What good---scientific---reasons buttressed this development? How rationally warranted was (and is) the adoption of cosmic inflation, if its methodological status seems so precarious? Helping oneself to a bit of melodramatic hyperbole, one may even worry: is the rationality of a major episode and research area in modern cosmology at stake? Small wonder therefore that cosmic inflation, and the ways in which it might be justified, received considerable philosophical attention. Extant work (e.g.\ \textcite{Smeenk:2017uof, McCoy2019-MCCEJA}) has attempted to ward off the spectre of irrationality by primarily emphasising plausible \textit{evidential} support of inflation (empirical as well as theoretical); the focus has tended to lie on (contemporary and historical) reasons to believe that cosmic inflation is true (or at least the best description of the early universe available)---in other words, on matters of \textit{confirmation}.

%Indeed, in a letter commenting on a sharp critique of inflation, \textcite{guth2017cosmic} state: ``according to the high-energy physics database INSPIRE, there are now more than 14,000 papers in the scientific literature, written by over 9,000 distinct scientists, that use the word `inflation' or `inflationary' in their titles or abstracts. By claiming that inflationary cosmology lies outside the scientific method [the critics] are dismissing the research of [...] a substantial contingent of the scientific community.'' WORDCOUNT

The present paper undertakes a complementary analysis. Rather than attempts to justify commitment to its truth, we’ll inspect rationales for \textit{working on} inflation---attempts to justify further research on it. In other words (viz.\ those of \textcite{Laudan1977-LAUPAI, Laudan1996-LAUBPA-3}), we’ll examine the \textit{pursuit-worthiness} of cosmic inflation---rather than its pretensions to acceptance/truth. Our analysis zeroes in on the scientific promise of cosmic inflation: why it plausibly does, and did, merit the extensive investigation it still enjoys---why it may be said to be (and to have been) rational to allocate substantial resources for its further study and elaboration.

The outline of the paper is as follows. \textbf{§\ref{inflation}} will review the basics of inflationary cosmology. \textbf{§\ref{pursuit}} will construe  past and present controversy over inflation in terms of historical and contemporary pursuit-worthiness (or promise) and acceptability (or confirmation). Concentrating on the former, we'll then sketch our philosophical strategy for cashing out pursuit-worthiness: to regard the display of salient theory virtues as indicators of promise. \textbf{§\ref{depth}} will begin with inflation's most important promise: better explanations---both with respect to fine-tuning issues of the inflation-less HBB model, and with respect to their ``depth''. \textbf{§\ref{unification}} will highlight inflation's unificatory power in its multifarious dimensions. The third principal virtue constituting inflation's promise is the ability to afford understanding, the subject of \textbf{§\ref{understanding}}. While the foregoing promises are more theoretical in nature, \textbf{§\ref{novelty}} will analyse a more empirical one: the prospects of testing inflation. In particular, we'll argue that inflation is distinguished from alternative proposals by predictive novelty. 

\section{Inflation}\label{inflation}

\noindent This section will first review the main ideas and foibles---certain fine-tuning issues---of the HBB-model (\textbf{§2.1}).\footnote{We follow standard presentations in \textcite{Baumann:2022mni, Weinberg:2008zzc, Mukhanov:2005sc, Dodelson:2003ft}, to which we refer for further details.} We’ll then (\textbf{§2.2}) expound the guiding principles of cosmic inflation; in particular, we’ll discuss how it relieves the HBB-model’s said fine-tuning features. Special attention will be paid to inflation’s ability to provide a plausible theory of the origins of structure formation in the universe (\textbf{§2.3}).

\subsection{The Hot Big Bang Model and its Discontents}

The classical Hot Big Bang (HBB) model is the standard cosmological framework for times as early as some fractions of a nano-second after the Big Bang. It applies our best theory of gravity, General Relativity (GR), to the entire (observable) cosmos. 

At its core, the HBB-model relies on GR's Friedmann-Lemaître-Robertson-Walker (FLRW) solution. The latter describes a homogeneous and isotropic universe on large scales, modelled as a perfect fluid (with energy density $\rho$ and pressure $p$). The resulting dynamics is given by the so-called Friedmann equations:\footnote{For convenience, we’ll use geometric units; in them $c=G=1$.}
\begin{equation}\label{velocity}
    H^2 \equiv \left(\frac{\dot{a}}{a}\right)^2 = \frac{1}{3}\rho - \frac{k^2}{a^2} + \frac{\Lambda}{3},
\end{equation}
\begin{equation}\label{acceleration}
    \dot{H}+ H^2 \equiv \frac{\ddot{a}}{a} = -\frac{1}{6}(\rho + 3p),
\end{equation}
where $a$ denotes the (time-dependent) scale factor (encoding the relation between physical distances and distances expressed in co-moving coordinates, and hence a measure of the expansion of space itself), $\rho$ the matter energy (comprising both matter and radiation), $p$ the pressure, $\Lambda$ the cosmological constant, $k$ the spatial curvature ($=-1, 0, +1$, corresponding to open, flat or closed universes respectively) and $H := \dot{a}/a$ the Hubble expansion rate (or ``Hubble parameter”). On the basis of current observations, the HBB-model’s matter content of the universe can be parsed into approximately 70\% Dark Energy (i.e.\ contributions due to the cosmological constant), 25\% Dark Matter, and 5\% ordinary (baryonic) matter. 

The Friedmann equations evolve the universe forwards (towards the ``cosmic dark ages”, where the universe becomes increasingly emptier and colder, see e.g.\ \parencite{Vaas2006}), as well as backwards towards an initial cosmic singularity, the Big Bang. For our present purposes, the backwards-evolution is of chief interest. According to the HBB-model, the universe began in a hot, dense state---what Peebles aptly monikered ``the primeval fireball”: all of matter contained within the observable universe was compressed within a miniscule fraction of the universe’s current volume; the concomitant, extremely high temperatures prevented atoms from forming in this primordial ``soup” of subatomic particles. During the subsequent expansion---within roughly the first three minutes after the Big Bang---the universe cooled down; the drop in temperature allowed atoms and elements to freeze out. Under the influence of gravity eventually clusters, galaxies and stars emerged; cosmic structure was born. 

Amongst the HBB-model’s successes, three stand out:

\begin{itemize}
\item The prediction of \textit{an expanding universe}---rather than a static one. This is most clearly manifested in the ``Hubble-Lemaître Law”, the linear redshift (recession velocity)-distance relation of far-away galaxies. 
\item The successful prediction of the \textit{cosmic microwave background (CMB)}, permeating the entire universe today, being nearly uniform in all directions, and exhibiting a nigh-perfect blackbody spectrum. The CMB represents a snapshot of relic radiation, originating from $380,000$ years after the Big Bang, when the universe transitioned from an opaque plasma-filled state to one where photons could travel freely through space.
\item The successful prediction and explanation of the abundances of light elements observed in the universe---\textit{primordial nucleosynthesis}: in the extremely hot cosmic furnace of the first three minutes, the light nuclei, primarily hydrogen and helium, were thus ``baked”. 
\end{itemize}
The triumphs of the HBB-model can’t be overestimated: not only did they transform cosmology from a more philosophical-speculative enterprise into an empirical-scientific discipline proper; they also instated the HBB-model as the nearly unanimous paradigm \parencite[Ch. 3\&4]{Kragh2013-KRACOC}. Soon, however, a fly in the ointment was spotted: the prerequisite initial conditions seem curiously fine-tuned. 

Observe first how the material constituents mentioned above add up: they exhibit an exquisite balance between gravitational attraction and expansion. It results in a spatially nearly flat universe: $\Omega_k =0.0007 \pm 0.0019$---where perfect flatness corresponds to $\Omega_k = 0$ \parencite{Planck:2018vyg}. Furthermore, at large scales, the distribution of matter and energy showcases remarkable homogeneity; the CMB engulfing us is found to have a uniform temperature of $2.73K$, with variations on the order of $10^{-5}$ \parencite{Hu:2002aa}.

Upon closer inspection, these oddities turn out to be even more eerie:
\begin{itemize}
\item The \textit{flatness problem} consists in the discomfort that the initial value of the density $\rho$ evokes. Not only is its presently observed value extraordinarily close to the critical density. Even if the universe happens to begin close to flatness, this initial state is dynamically unstable.\footnote{In the terminology of dynamical systems analysis, it's a bifurcation point.} We can see this directly from the curvature parameter $\Omega_k$:
\begin{equation}
    \Omega_k := 1 - \Omega = \frac{k}{a^2H^2},
\end{equation}
where $\Omega$ is the ratio of the matter density $\rho$ to the critical density $\rho_c$. In a dynamically evolving universe, $\Omega_k$ will rapidly diverge from zero as the scale factor $a$ and Hubble parameter $H$ evolve over time (i.e.\ $a(t) \propto t^{1/2}$ during radiation domination, and $a(t) \propto t^{2/3}$ during matter domination). To obtain the degree of spatial flatness we observe in the present epoch, $\rho/\rho_c \approx 10^{-16}$ during Big Bang nucleosynthesis---and, if we extrapolate back to a GUT or Planck scale, even $\rho/\rho_c \approx 10^{-55}$\parencite[p.25]{Baumann2009}! The initial conditions must be uncannily tuned to match observations. 
\item The universe is homogeneous over vast, causally disconnected regions of space. Indeed, we can divide the observable universe into $\sim 10^{84}$ causally disconnected patches that display nearly the same temperature, with only tiny deviations \parencite[p.227]{Mukhanov:2005sc}. This seems inexplicable: no known causal processes exist that could have produced such uniformity. Hence, such a baffling degree of uniformity in the matter-energy sector must be put in by hand. This is the so-called \textit{horizon problem}.

\end{itemize}

\subsection{Basics of Inflation}

Inflation amends the FLRW dynamics of the HBB-model for the time before roughly one second: it postulates a preceding period of rapid, exponential expansion (smoothly succeeded by the HBB-model’s description). On the one hand, all the key successes of the HBB-model are thereby left in tact; on the other hand, the growth spurt during that inflationary period overcomes the HBB-model’s fine-tuning issues. 

In its most basic incarnation, inflation involves a scalar field $\varphi$, the inflaton; it’s supposed to move in the potential $V(\varphi)$. Plugged into the Friedmann equations as a matter source, it drives the expansion. 
More specifically, according to inflation, the early universe had initially been in a metastable state, the so-called false vacuum--- a non-zero vacuum state that persists for an extended period before it decays into a more stable state.

For a while, the inflaton is trapped in that false vacuum at a high potential energy. It produces a nearly constant energy density that dominates over its kinetic energy.\footnote{According to the so-called strong energy condition, a field’s energy density of any form of matter must be non-negative and shouldn’t exceed its pressure. It’s usually taken to be a constraint to which ordinary matter conforms (see \textcite{Curiel:2014zba} for a critical review). The inflaton, by contradistinction, counts as exotic in that it violates the strong energy condition. During its ``slow-roll” through the potential (during which $\dot{\varphi}^2\ll V$), we get for its equation of state: 
\begin{equation}
    w := \frac{p}{\rho} = \frac{\frac{\dot{\varphi}^2}{2} - V(\varphi)}{\frac{\dot{\varphi}^2}{2} + V(\varphi)}\approx -1.
\end{equation}
} Inflation occurs while the inflaton “slowly rolls down” the potential---if (and as long as) the kinetic energy density only makes a
small contribution to the total energy density; the latter accordingly stays roughly constant. This creates a repulsive gravitational effect, mimicking that of a cosmological constant: the inflaton’s slow-roll induces a phase of approximately exponential (``quasi-de Sitter”) expansion for the universe’s scale factor, $a(t) \propto e^{Ht}$. The dynamical effect of that expansion provides a straightforward resolution to the flatness and horizon problems. 

Consider the curvature parameterisation introduced earlier, $\Omega_k = k/(aH)^2$. During inflation, the scale factor grows exponentially while the Hubble constant is approximately constant. The curvature parameter $\Omega_k$ is thus is driven towards zero (and the density parameter $\Omega$ is driven towards unity) \parencite[p.8]{Guth:2004tw}. In other words, inflationary dynamics naturally induce a flat geometry. Given a sufficient period of inflation, this holds for almost any initial value of the curvature at which one starts. 

Inflation resolves the horizon problem by establishing past causal connectivity among distant regions. During inflation, the observable universe \textit{was} causally connected (when it was much smaller). However, the subsequent exponential expansion of space moved these regions outside of each other's horizons, creating the appearance of present-day causal \textit{dis}connection. But inflation not only establishes causal connectivity. It also explains the observed homogeneity: any initial inhomogeneities are rapidly inflated away, leading to a highly homogeneous universe \parencite{Brandenberger:2016uzh, East:2015ggf}.

What happened once inflation came to an end?\footnote{Exactly \textit{how}---or even to what extent---inflation comes to an end, is a topic of on-going research outside of the present paper’s ambit. We refer to the literature for further discussion.} The false vacuum isn’t stable; the inflaton gradually evolves towards its true vacuum state. As it slowly rolls down its potential, it transitions from the false to the true vacuum (i.e.\ the global energy minimum)---a process known as ``reheating”. It releases large amounts of energy. The energy is efficiently converted into radiation and matter, producing the standard particles (assumed in the HBB-model ab initio), as well as primordial density perturbations. The latter constitute one of inflation’s most significant achievements; we’ll briefly comment on it next.

\subsection{Generating Cosmic Structure}
Prior to the advent of inflation, the understanding of structure formation in the universe was limited to a phenomenological description, devoid of deeper physical principles: the right initial perturbations, their power spectrum and amplitude had to be prescribed by hand in order to reproduce the observed large-scale features (see \textcite{Smeenk2018-SMEIAT} for historical details). 

Inflation, by contrast, provides a theory that can account for the origins for these perturbations through the application of standard\footnote{A point forcefully stressed by \textcite{Wallace:2021}.} quantum field theoretic techniques to the classical inflaton. ``Rather than pulling the initial spectrum out of a hat, as one might suspect of the earlier proposals, the inflationary theorist can pull a [nearly scale-invariant] spectrum [...] out of the vacuum fluctuations of a quantum field'' \parencite[p.9]{Smeenk2018-SMEIAT}.

The quantum fluctuations in the inflaton field generate the density perturbations, which serve as the seeds for the formation of cosmic structures, including galaxies, clusters, and stars. These fluctuations arise from minuscule quantum variations in the value of the inflaton field during the inflationary epoch; the exponential expansion stretches the (sub-)microscopic fluctuations to macroscopic scales, where subsequent gravitational collapse leads to the formation of cosmic structure.

According to inflation, these perturbations---whose imprint on the CMB can, and has been, empirically probed---possess distinct statistical properties:
\begin{itemize}
\item \textit{An adiabatic spectrum}: the power spectrum of density fluctuations is independent of the type of matter/energy involved.
\item \textit{Gaussian distribution}: the fluctuations are normally distributed; their Fourier modes are uncorrelated.
\item \textit{Nearly scale-invariant spectrum:} the power spectrum of density fluctuations exhibits approximate, but not exact, independence from length scales.\footnote{The spectral index $n_s$ quantifies the scale-dependence of the power spectrum. With the so-called ``slow roll parameters” $\varepsilon:=-\frac{\dot{H}}{H^2}$ and $\kappa:=\frac{\dot{\varepsilon}}{H\varepsilon }$, both of which are related to the derivatives of the potential, one gets: $n_s - 1 =- 2\varepsilon - \kappa$. Inflation occurs for the slow-roll approximation, i.e.\ $|\kappa|, \varepsilon \ll 1$. Accordingly, $n_s – 1 \approx_{\neq}0$. Consequently, the power spectrum deviates \textit{slightly} from scale-invariance ($n_s \simeq 1$).}
\end{itemize}
Current measurements are in excellent agreement with these predictions \parencite{Planck:2018vyg}.

\section{Past and Present Controversy Over Cosmic Inflation---Historical and Contemporary Pursuit-Worthiness}\label{pursuit}

\noindent The historical challenge of inflation, as we saw above, consists in justifying its ascendancy to mainstream research despite scant evidence. Here, we'll circumscribe our paper's overall agenda, and philosophical strategy for tackling that challenge. Our project also bears on an on-going, recent debate in primordial cosmology between advocates and detractors of inflation over its scientific status (\textbf{§\ref{strife}}). \textbf{§\ref{acceptpursuit}} will highlight an aspect of this debate that has received insufficient attention---the issue likewise at the heart of the historical challenge: the distinction between reasons for accepting a theory, and reasons for pursuing it. To advance the debate both with respect to the historical challenge as well as with respect to the recent controversy, we propose to defend inflation's historical and continuing pursuit-worthiness via certain theory virtues; they serve, we submit, as plausible indicators of pursuit-worthiness (\textbf{§\ref{rational}}).

\subsection{Strife Over Inflation} \label{strife}
Cosmic inflation polarises, with perplexing intensity, the cosmology community. On the one hand, inflation is nigh-universally embraced as ``a broadly accepted cosmological paradigm'' \parencite[p.62]{Linde:2014nna}. The PLANCK collaboration, the most recent experiment to probe the CMB, repeatedly emphasise that their data is ``remarkably consistent with a spatially-flat $\Lambda$CDM cosmology with purely adiabatic, Gaussian initial fluctuations, as predicted in simple inflationary models'' \parencite[p.2]{Planck:2018vyg} and provides ``very strong support for the inflationary paradigm'' \parencite[p.30]{Planck:2018nkj}. Similarly, following the successful WMAP and PLANCK missions, \textcite[p.1]{Guth:2013sya} proclaim: ``these generic predictions are consequences of simple inflationary models [...]. To date, every single one of these inflation-scale predictions has been confirmed to good precision.'' \textcite[p.1]{Chowdhury:2019otk} conclude: ``it seems fair to say that inflation is viewed as the best paradigm for the early universe by a vast majority of scientists working in the field of cosmology''. 

On the other hand, a minority group of physicists---amongst them, one of inflation’s co-inventors, Paul Steinhardt, as well as other prominent figures such as Roger Penrose or Robert Wald---gainsay the reigning paradigm. They scathe inflation for a battery of reasons. Some query its evidential situation: \textcite{Ijjas:2015hcc, Ijjas:2013vea}, for instance, avow that the continued failure to detect primordial gravitational waves is disconcerting; these observations strongly disfavour the inflationary paradigm. Other issues they table concern conceptual conundrums. They chiefly revolve around prerequisite initial conditions for inflation \parencite{Penrose:1988mg, Hollands:2002yb}\footnote{They argue that the initial state we should expect to emerge from a gravitational singularity would make inflation overwhelmingly \textit{unlikely} to occur.} and the so-called trans-Planckian problem \parencite{Martin:2000xs}\footnote{That is, the technical and methodological problems surrounding the conceivable sensitivity of cosmological observables to quantum-gravitational physics at the Planck scale, ``a regime where these theories are known to break down” \parencite[p.1]{Martin:2000xs}.}. But also broader methodological misgivings are voiced. They target especially multiverse scenarios (and the status of fine-tuning, probabilities and predictions in general), as implied by some inflationary models (see e.g.\ \textcite{Guth:2007ng}).

In the foregoing sketch of the  ``cosmic controversy'' we can descry two interwoven themes. To advance the debate, it will prove useful to disentangle them:
\begin{itemize}
    \item \textit{Evidential-Epistemic Status (EES)}: Has inflation been confirmed? How compelling is the evidence for it? Is it rational that ``cosmologists appear to accept at face value […] that we must believe the inflationary theory because it offers the only simple explanation of the observed features of the universe” (Ijjas et al., 2017, p.34)?
    \item \textit{Promise \& Potential (P\& P)}: Does inflation deserve the attention and research efforts that it enjoys as the ``dominant paradigm”? Is inflation (still) promising enough to (continue)---does its potential augur sufficiently well---to justify these investigative investments? 

\end{itemize}
At first blush, the issues under \textit{(EES)} seem preponderant in the debate. This is to be expected: questions of confirmation, proof, likeliness-to-be-true, etc.\ have traditionally preoccupied scientists and philosophers of science (see e.g.\ \textcite{Smeenk:2017uof, Azhar:2016mpg, DawidManuscript-DAWIIC}). Yet, the questions under \textit{(P\&P)} are no less present. Arguably they are even of greater \textit{practical} relevance for scientists: they are tied up with the direction and organisation of \textit{future} research (see also e.g.\ \textcite[Sect.7]{Wolf:2022yvd}).

To begin with, \textcite{guth2017cosmic} acknowledge, that ``(i)nflation is not a unique theory but rather a class of models based on similar principles”. In the same spirit, inflation is recurrently (e.g.\ \textcite{Guth:2013sya, Guth:2005zr}) referred to as a ``framework”. As such, questions of evidential/empirical warrant become somewhat awkward, if not moot (cf.\ \textcite{curiel2021framework}).\footnote{\textcite[p.18]{Scott:2018adl} indeed writes: ``[inflation] is undoubtedly an appealing idea, and there is a great deal of circumstantial evidence to support it---so I think it’s entirely reasonable to be a \textit{fan} of inflationary cosmology. But since inflation is really a framework rather than a model, we can’t assert that any of the observations actually prove that inflation is correct.''} Frameworks are more plausibly assessed in terms of appropriateness/viability, utility or fertility in the quest for more refined, concrete models of empirical phenomena.\footnote{\textcite{guth2017cosmic} are plausibly read in this sense when they write: ``(n)o one claims that inflation has become certain; scientific theories don’t get proved the way mathematical theorems do, but as time passes, the successful ones become better and better established by improved experimental tests and theoretical advances.”} The pertinent question for a framework is instead: should we commit to deploying it for further investigations? The situation, according to \textcite{guth2017cosmic}, ``is very similar to the early steps in the development of the standard model of particle physics, when a variety of quantum field theory models were explored in search of one that fit all the experiments”.\footnote{This historical parallel (whose accuracy we won't question) is grist to our mills, and one source of inspiration for us, as far as our paper's agenda---a defence of inflation's pursuit-worthines via super-empirical theory virtues (\textbf{§3.3})---is concerned, see \textcite{Schindler2014-SCHAMO-12, RuizdeOlano2023-RUICOP-2}.} In an earlier article, \textcite[p.113]{Guth:2013sya} (our emphases) likewise stress that ``the success that inflation has had in explaining the observed features of the universe give us \textit{motivation to explore} the speculative implications of inflation for questions far beyond what we can observe”. 

By the same token, critics of inflation are also frequently less concerned with evidential warrant than one might initially think. Summoning ``new reasons to consider competing ideas about the origin and evolution of the universe”, \textcite{ijjas2017pop} rally for the exploration of alternatives to inflation---clearly in the spirit of \textit{(P\&P)} rather than \textit{(EES)}. The conceptual anomalies and foundational problems that inflation’s critics have discriminated are of course germane to an answer to \textit{(EES)} or \textit{(P\&P)}: should we nonchalantly shelve them as open questions and challenges for future developments---Kuhnian puzzles\footnote{\textcite{steinhardt2022disowns}'s complaint that ``(m)ost astrophysicists have gone about their business testing the predictions of textbook inflationary theory without worrying about these deeper issues, hoping that they would eventually be resolved” indeed sounds like a paraphrase of Kuhnian normal science: anomalies and foundational questions are shoved aside in order to further advance the paradigm’s disciplinary matrix. The usage of Kuhnian terminology in the physics literature is, however, loose.} that, in due course, will be solved? Or are they so severe as to undercut inflation’s pretences to confirmation (as per \textit{(EES)}), or so damning that they mar the theory’s potential (as per \textit{(P\&P)})? \textcite{Ijjas:2014nta} rhetorically ask: ``is it time to seek an alternative cosmological paradigm?”. The problems that they diagnose, in their eyes, signal the bankruptcy of inflation as a research programme: it ought to be ``abandoned” \parencite{ijjas2017pop}---in the sense of \textit{(P\&P)}. While not impugning that important questions remain, \textcite[p.118]{Guth:2013sya} by contrast insist that ``the inflationary paradigm, with its many successes, provides a framework within which such additional questions may be pursued''. Might philosophy of science have something to contribute to this debate?

\subsection{Acceptance \textit{vs.} Pursuit}\label{acceptpursuit}

The above rubrics \textit{(P\&P)} and \textit{(EES)} track a familiar distinction in the philosophy of science literature. \textcite[p.111]{Laudan1996-LAUBPA-3} (see also \textcite{Whitt} or \textcite{Barseghyan2017-BARHCA-4}) has ``proposed distinguishing sharply between the rules of appraisal governing acceptance and the much weaker and more permissive rules or constraints that should govern ‘pursuit’, noting that ``there is a whole spectrum of cognitive stances that scientists can adopt toward their theories''. 

Each prompts a different epistemological agenda, i.e.\ normative evaluations of rational warrant. Justification of pursuit, in particular, should be demarcated from justification of acceptance. Within the ``context of \textit{pursuit}'', one inquires into whether a theory deserves further development: ``(t)o consider a theory worthy of pursuit amounts to believing that it is reasonable to work on its elaboration, on applying it to other relevant phenomena, on reformulating some of its tenets'' \parencite[p.3]{Barseghyan2017-BARHCA-4}. By contradistinction, acceptance concerns the more traditional objective of epistemology: evidential warrant for, or confirmation of, a theory.

The distinction between pursuit and acceptance is profitably applied to the case for inflation, both with respect to its historical and present challenge. \textcite{guth2017cosmic} declare: ``there is no disputing the fact that inflation has become the dominant paradigm in cosmology''. In light of Laudan's distinction, this admits of two (mutually non-exclusive) readings:
\begin{itemize}
    \item \textit{(P)} Inflation has become the dominant paradigm \textit{pursued} in contemporary cosmology.
    \item \textit{(A)} It has become the dominant paradigm \textit{accepted} in contemporary cosmology.
\end{itemize}
The focus of inflation's historical challenge lies on \textit{(P)}. Its descriptive accuracy---as a sociological claim---won't be called into question. We'll bracket issues related to the support or degree of confirmation cosmic inflation has---or maybe has not---received. Instead, we'll take up the philosophical/normative challenge that \textit{(P)} provokes: what can and could justify inflation's pursuit? Does inflation merit its privileged status as a theory insofar as scientists decide to work on it? In short, our subsequent goal will be to tackle \textit{(P\&P)}: we'll reconstruct the rationales that undergird inflation's historical and contemporary pursuit-worthiness. 

Given the distinctness of those two ``cognitive stances'', it's unsurprising that rationales for pursuit and acceptance needn’t coincide: what counts as a good reason to pursue a theory may not count as a good one to accept it to be true (and vice versa). The existing literature on inflation has largely neglected the question of pursuit-worthiness: either the distinction hasn't been broached at all---or the discussions have focused on evidential criteria for acceptability (see e.g.\ \textcite{Smeenk:2017uof, Earman1999-EARACL}).\footnote{\textcite[p.753]{penrose2004road} is a representative example of the first case: he smoothly segues from ``Are the motivations for inflation valid?'' to ``What reason is there to believe that such an inflationary picture of the universe is likely to be close to the truth?''. \textcite[p.3]{Earman1999-EARACL} illustrate the second case: ``the idea of inflation is, of course, most interesting and worth pursuing”. Regrettably, they don't spell out \textit{what} makes inflation so ``interesting and worth pursuing''; instead, they proceed with casting doubt on its \textit{evidential} status.} The omission is lamentable in at least three regards:
\begin{itemize}
    \item The case for inflation’s acceptance, found in the philosophical literature to-date---the most sustained one invoking Dawid’s Meta-Empirical Theory Assessment (see \textcite{Dawid2013-DAWSTA, DawidManuscript-DAWIIC, McCoy2021-MCCMSF-2}; cf.\ \textcite{WolfMeta} for a response)---is predicated on controversial methodological premises.\footnote{\textcite{Cabrera2021-CABSTN} has explicitly argued that Dawid's Meta-Empirical Theory Assessment is more plausibly understood as warranting pursuit rather than, contrary to Dawid's intentions, underwriting a form of meta-empirical confirmation.} A positive methodological appraisal, independent of such assumptions, would hence be of substantial interest to those unnerved by a persistent and pronounced misalignment of philosophical judgement and scientific practice (see e.g.\ \textcite{Schindler2014-SCHAMO-12} and \textcite{McCoy2019-MCCEJA}). An evaluation of the theory’s pursuit-worthiness provides such an appraisal.
    \item Even if successful, any analysis of inflation’s credibility as an account of the early universe inevitably fails to address the historical challenge: it seems doubtful that \textit{any} of the relevant empirical or theoretical arguments to which today an advocate of inflation might plausibly point, were available when cosmologists started merrily embracing inflationary cosmology. Are we forced, then, to relegate the adoption of inflation to a chapter of what \textcite[p.91]{Lakatos1978-LAKTMO} dubs ``external history'', sociological happenstance (or just a lucky guess), ultimately defying rationality---``a matter of mob psychology”?\footnote{This is indeed what \textcite[Sect.2]{Earman1999-EARACL} seem to suggest.} A methodological assessment of pursuit-worthiness allows us to forgo that conclusion (without denying the role of non-epistemic, i.e.\ social and institutional, factors in the actual historical reception of inflation).
    \item  We think that the thrust of such an assessment does greater justice to the prevalent attitudes we discern in the cosmology community. Admittedly without having conducted a proper poll, interviews, etc.\, our impression is that most physicists working on inflationary cosmology are primarily interested in questions of pursuit.\footnote{We’d like to stress that pursuit-worthiness as an object of normative evaluation is also significantly more congenial to pluralism than a theory’s evidential/confirmational credentials: typically, the simultaneous exploration of alternative theories, at least as minority research programmes, is---for quite general reasons (see e.g.\ \textcite{Bschir2015-BSCFAP, Shaw2022-SHAOTV-2, Lohse2020-LOHTCP})---salubrious to science; as such, challenges to orthodoxy should be welcomed. This matches indeed our impression from the cosmology community: while inflation is pursued predominantly (as we’ll argue: for good reasons!), deviant viewpoints are taken seriously---but have so-far failed to convince the mainstream of their superior pursuit-worthiness. 
    
    In this regard, a snippet from a conversation on the ``cosmic controversy” of (\textbf{§3.1}) is instructive: ``Loeb does not necessarily think that inflation is wrong. But he thinks alternatives should be taken more seriously. ‘I have no disagreement with that’, says Guth. ‘[…] And I would also disagree with claims that any of the current alternatives to inflation have a comparable stature” \parencite{Chown} (see also \textcite[p.243]{guth1997thesis}).} 

\end{itemize}

\subsection{The Rationality of Pursuit: Virtues as Indicators of Promise}\label{rational}

But how to justify a theory's pursuit? What, in particular, constitutes pursuit-worthiness of inflation, legitimating the concerted research expenditures that have been, and continue to be, devoted to its further investigation?  

It would be overambitious for this paper to enunciate, let alone defend, a definitive catalogue of criteria for pursuit-worthiness (see e.g.\ \textcite{Seselja2014-EELEJI, Lichtenstein2021-LICMSD, Fleisher2022-FLEPAI-5, Shaw2022-SHAOTV-2}). Instead, our philosophical working hypothesis will take its cue from \textcite{Nyrup2015-NYRHER}'s reading of Peircean abduction (see also \textcite{McKaughan}): we’ll presume that certain prima facie attractive features of inflation as a theory, so-called ``theory virtues'' (see e.g.\ \textcite{IvanovaForth, Keas2018-KEASTT-2, mcmullin2014virtues, Kuhn1977-KUHOVJ}), provide  good reasons to further explore it. Our strategy will be to regard those theory virtues as indicators of promise; such virtuous theories merit further pursuit.\footnote{The analysis in \textcite{McCoy2015-MCCDIS} and \textcite{McCoy2019-MCCEJA} can also be read as being concerned with pursuit-worthiness. In contradistinction to our project, though, McCoy largely follows \textcite{Laudan1977-LAUPAI} in taking progress with respect to problem-solving as a criterion for pursuit-worthiness (see also \textcite{McCoy2023-MCCUTP-2}). In particular, McCoy investigates whether inflation can be said to solve certain (primarily conceptual) problems that its competitor, standard General Relativity with special conditions, fails to solve satisfactorily (see \textbf{§4.1)}. McCoy arrives at a negative response. Our analysis, with its focus on explanatory virtues (and theory virtues more generally) as indicators of pursuit-worthiness, is intended as complementary.} The guiding thought is analogous to a cost-benefit analysis (where the potential benefit is an empirically successful and cognitively highly  valuable theory, and the costs correspond to the energy and time of the resources invested into the theory, should in the end the theory have to be jettisoned): theories that exhibit certain virtues hold such promise---pledge such desirable accomplishments with respect to the aims or values of science (see e.g.\ \textcite[Ch.2]{Nola2007-NOLTOS-3}, first and foremost ``to find satisfactory explanations of whatever strikes us as being in need of explanation” \parencite[p.132]{Popper1983-POPRAT}---that they are worth pursuing, even if the \textit{ultimate} successes of that endeavour---to produce empirically corroborated theories---remain uncertain. That is, virtuous theories \textit{would} be so scientifically and epistemically valuable, \textit{if} they \textit{were} true, that it seems rational to invest resources into figuring out whether they \textit{actually} are true---of course, at the risk of such hopes coming to naught.
In particular, the adopted strategy implies that, while not necessarily licensing an inference to truth (or likelihood to be true), a theory’s explanatory ``loveliness” \parencite{Lipton1991-LIPITT}---the display of certain virtues \textit{in the explanations} it underwrites---can justify its pursuit.\footnote{Appraisal of pursuit-worthiness can set aside considerations of truth-values (and realism in particular). As \textcite[p.10]{Cabrera2022} writes, ``(b)y contrast, proponents of IBE [inference to the best explanation---such as Lipton]” typically claim that a strong, successful inference to the best explanation can rationally justify full-fledged belief in a hypothesis. In other words, it is standard to regard IBE as belonging to the context of justification, rather than the context of discovery or context of pursuit." The explanatory loveliness at the core of \textbf {§4} is, in contrast to Lipton's original usage, thus as it were an inference to the---ceteris paribus---most pursuit-worthy explanation. We thank an anonymous referee for pressing us on this.}
With this philosophical strategy we'll kill two birds with one stone---both the historical question of inflation's pursuit-worthiness, as well its continued pursuit-worthiness.

The task before us is thus twofold. First, the virtues that cosmic inflation exhibits will have to be made explicit. We'll hone in on various dimensions of explanatory depth (\textbf{§4}), unificatory power (\textbf{§5}), understanding (\textbf{§6}), and novelty (\textbf{§7}). 
Secondly, in order to vindicate the rationality of adopting cosmic inflation, it's incumbent on us to render plausible a connection between those virtues and pursuit-worthiness. Our arguments (further elaborated on a rolling basis in the subsequent sections) pivot on classic epistemological \textgreek{ἀρεταί}:
\begin{itemize}
    \item \textit{Pursuit and better explanations:} While explanations are  universally coveted in science, deep explanations (in senses to be explicated) are especially  attractive: they reduce ad-hocness, and allow epistemically cautious (or moderately conservative) theoretical innovation.
   \item \textit{Pursuit and unification:} A theory's broad and/or diverse range of applicability is an intuitive desideratum, universally cherished. Unifying power, more broadly construed, ensures a theory's coherence, both with respect to its organic, internal structure, and with respect to our background knowledge. 
   \item \textit{Pursuit and understanding:} Some theories enhance (the quality of) our understanding more than others. With understanding being a key cognitive goal of science, it follows that, ceteris paribus, the more a theory is conducive to that goal, the more desirable it is.
    \item \textit{Pursuit and (novel) predictions:} Often\footnote{Although not exclusively: think of the pursuit of toy models (see e.g.\ \textcite{WattMisner}.)}, scientists pursue a theory, hoping that (one day) it might accrue sufficient empirical support to be accepted. One then naturally demands that the theory-to-be-pursued cross some minimal plausibility threshold: not only should it allow for at least rough-and-ready tests that inform researchers that the theory isn't completely off track; the theory should also display minimal---to be sure: tentative---positive indicators that it's roughly on the right track. This is what successful novel predictions can achieve.
\end{itemize}.

\section{\textit{Promise 1:} Better Explanations}\label{depth}

\noindent Typically, introductions to inflation hail its capacity for explaining ``many features of the observable universe lacking an explanation in the standard big-bang model'' \parencite[p.127]{guth1984inflationary}. It therefore seems fitting to commence our case for inflation’s pursuit-worthiness by scrutinising its promise of better explanations. What renders them \textit{better} than those of the inflation-less HBB-model? First, inflation offers advantages with respect to fine-tuning issues that the HBB-model faces (\textbf{§4.1}). Secondly, inflationary cosmology purveys explanations that are deeper than those of the HBB-model (\textbf{§4.2}-\textbf{§4.3}).

\subsection{Fine-Tuning}\label{Relief for Fine-tuning}
Intuitions about fine-tuning exert tremendous influence on cosmology (see e.g.\ \textcite{Friederich2020-FRIMTA-2}), both generally (usually in the context of alleged fine-tuning for life), as well as specifically with respect to inflation (see e.g.\ \textcite{guth1984inflationary}). We'll concentrate on the latter context. Despite its popularity, it has proven difficult to spell out, let alone uncontroversially establish, the meaning and significance of fine-tuning (see \textcite{McCoy2015-MCCDIS} for a careful analysis). The most common attempts involve probabilistic reasoning, appeals to symmetry principles, or discomfort with large numbers. They require, at best, considerable further developments, and hinge on controversial premises (see also \textcite{Hossenfelder:2018ikr}). Probabilistic arguments in particular, by dint of which cosmologists frequently try to make plausible statements that initial conditions in the HBB-model are ``unlikely”, are notoriously tenuous: at best they lack a cogent justification; at worst, they are hopelessly misguided stabs at the problem \parencite{Schiffrin:2012zf, Curiel:2015oea, McCoy2018-MCCTII-2}.

One may even demur that the flatness and horizon problems aren't ``problems'' at all. Generally, when we evaluate the success of a dynamical model, ``the choice of initial conditions is usually made only in order to facilitate the comparison” between the equations of motion and the evolution of the system. ``(T)he equations [of the HBB-model], with suitably chosen initial conditions, do a perfectly good job of describing the evolution of the universe” \parencite[p.6]{Albrecht:2000gh}. Following this logic, nothing seems particularly enigmatic about the initial conditions that the HBB-model requires for its empirical adequacy; they are the brute facts of the best-fitting model. Rather than an ad-hoc dodge, this inference pattern tallies with standard practice (especially in the historical sciences, such as geology or paleontology). By the same token, it’s not obvious that the observed homogeneity of the universe portends any deeper mystery.  ``The horizon problem is not a failure of the standard big bang theory in the strict sense, since it is neither an internal contradiction nor an inconsistency between observation and theory. The uniformity of the observed universe is built into the theory by postulating that the universe began in a state of uniformity'' \parencite[p.184]{guth1997inflationary}. 

Intuitions about fine-tuning---and inflation’s improvements on the HBB-model in this respect---may be preserved, whilst evading several of the just-mentioned obstacles, if one construes fine-tuning in terms of instantiations of \textit{extraordinary types}---types that, given our background knowledge, we don’t expect. We’ll take our cue from \textcite{baras-ms} (see also \textcite{Baras2020-BARCFE-4}). He ponders when certain facts or phenomena ``call” for an explanation. For our purposes, we’ll take this to translate into: which facts are sufficiently ``suggestive” or ``peculiar” so as to make an explanation \textit{desirable} (but neither compulsory nor guaranteed to exist)?

According to Baras, a particular fact or phenomenon $x$ calls for an explanation if it instantiates an extraordinary type. A type is unusual or striking in light of what we know, or even have reason (not) to expect. Our background knowledge about the pertinent domain determines the definition/individuation of those types, as well as their classification as \textit{extraordinary}.\footnote{Consider, for illustration, Boltzmannian Statistical Mechanics (in its typicality interpretation, see e.g.\ \textcite{Lazarovici2015-LAZTIA}). The Past Hypothesis---the positing of initial conditions of a system, sufficiently large to contain our observable universe, that give rise to a so-called Thermodynamic Arrow of Time---doesn’t call for an explanation, according to Baras’ proposal. Types of micro-configurations are naturally given by the macro-states that they realise; these types are characterised via their thermodynamical properties. Furthermore, the space of initial conditions is endowed with a natural measure that allows us to ``count” any measurable set of initial conditions. It turns out that relative to this measure \textit{most} initial conditions entail that the system exhibits a Thermodynamic Arrow of Time---that its entropy never decreases. The Past Hypothesis, on this reasoning, doesn’t involve an instance of an extraordinary type; it doesn’t ``call” for an explanation.} The occurrence of ordinary/non-striking types of entities \textit{coheres} well (has strong and plentiful ``inferential links'') with background knowledge (including the natural resources of the theory ``in charge” of the pertinent domain): our background knowledge provides good reasons to expect those types---rather than others (see \textcite{Schindler2018-SCHACC-4} and \textcite[Ch.5]{Schindler2018-SCHTVI-5}).\footnote{Following \textcite{Schindler2018-SCHACC-4} and \textcite[Sect.4.1]{duerr-wolf-forthcoming}, one may conversely take the instantiation of extraordinary/striking types to signal a theory’s ad-hocness. Recall also the traditional link between explanation and expectation stressed by e.g.\ Hempel or Salmon (see \textcite[Sect.4]{sep-scientific-explanation}).} Coherence is a routinely discussed theory virtue; its epistemic clout is widely recognised (see e.g.\ \textcite{BonJour1985-BONTSO-4, Keas2018-KEASTT-2}). In line with our rationale for pursuit-worthiness, we’ll treat coherence as an—of course defeasible—indicator of pursuit-worthiness (see also \textcite{Seselja2014-EELEJI} for a similar idea): theories that alleviate fine-tuning issues are pursuit-worthy as they promise to enhance the coherence of our knowledge. 

NB: Naturally defined measures allow a categorisation of extraordinary types. Thanks to the appeal to coherence, however---an explanatory relation that admits of different strengths and kinds---typicality or probability measures \textit{aren’t necessary} for the classification of types as extraordinary. Extraordinariness is commonly, and can be, adjudicated on qualitative grounds. 

We thus arrive at the following proposal for a modest interpretation of fine-tuning issues as \textit{tentative} inklings, a wish list, for pursuit-worthy theories:
\begin{itemize}
    \item \textit{FT-1}: A fact (e.g.\ initial conditions) is fine-tuned, iff it instantiates what, given our background knowledge, counts as an extraordinary type—a type not optimally cohering with our warranted expectations. 
    \item \textit{FT-2}: Ceteris paribus---in the absence of more
glaring empirical or theoretical anomalies---a theory that explains a fine-tuned fact deserves pursuit more than one that doesn’t explain it.

\end{itemize}
This deflationary reading of fine-tuning as suggestive oddities that can marshal theory-pursuit jibes with physical practice: ``(t)he reception of Guth’s case for inflation [...] reflects, in part, a common strategy: using fine-tuning as a \textit{guide to developing new theories}” \parencite[p.218, our emphasis]{Smeenk2018-SMEIAT}. Our proposed reading, FT-1 \& FT-2, fully concurs with Carroll's circumspect stance towards fine-tuning: ``(a)ny fine-tuning is necessarily a statement about one’s expectations about what would seem natural or non-tuned.” The point of fine-tuning arguments is ``to look for \textit{clues} in the state of the universe that might help \textit{guide us towards} a more comprehensive theory'' \parencite[p.4, our emphasis]{Carroll:2014uoa}.

Baras' account\footnote{Baras also applies his proposal to fine-tuning. He arrives at a negative verdict: it doesn't, he argues, call for an explanation. The fine-tuning Baras examines, is however fine-tuning of the universe \textit{for life}. In contrast to the latter, \textit{the HBB-model's} fine-tuning issues are epistemically well-understood, and draw on the theory’s own natural resources (such as GR's dynamics).} vindicates the hunch that, without inflation, the HBB-model faces three principal fine-tuning feature that call for an explanation in the foregoing sense (recall \textbf{§\ref{inflation}}):
\begin{itemize}
    \item \textit{Flatness problem:} Suppose that we pare down the types of universes to those with a homogeneous and isotropic large-scale structure (i.e.\ FLRW models). They are naturally classified geometrically, i.e.\ according to their spatial curvature---describing either an open, closed or flat universe. Flat universes, such as ours are extraordinary in light of GR's dynamics: they are unstable/fragile \parencite[Sect.3]{McCoy2020-MCCSIC}. That is, GR’s dynamics governing the universe’s evolution rapidly amplifies ever-so-slight initial deviations from flatness at any point in time; the universe would be driven towards either openness or closedness. 
    \item \textit{Horizon problem:} Types of universes as uniform (homogeneous and isotropic) as ours are extraordinary given that most of the visible universe hasn't been in causal contact, if we adopt the HBB-model's dynamics. Our background knowledge has us \textit{not} expect causally disconnected regions to share the same properties \parencite[p.23]{Baumann2009}.
    \item \textit{Superhorizon correlations:} The universe isn't only flat and homogeneous. But the density fluctuations---the statistical \textit{deviations} from perfect homogeneity---that fill the universe ``are correlated over apparently acausal distances. This [...] begs for a dynamical explanation'' \parencite[p.138]{Baumann:2022mni}. 
\end{itemize}
Inflation answers those ``calls for explanation''. As reported in \textbf{§\ref{inflation}}, it demotes the above facts to tokens of \textit{ordinary} types, rather than \textit{extraordinary} types. Given inflation, uniformity, spatial flatness, and superhorizon correlations are to be \textit{expected} in our universe; these features become ``generic predictions”. ``(F)rom almost any initial conditions the universe evolves to precisely the state that had to be assumed as the initial one in the standard model'' \parencite[p.116]{guth1984inflationary}.

Of course, inflation can’t achieve these feats with an \textit{arbitrary} dynamics; its dynamics must in \textit{some} sense be special. At first blush, we trade-off prima facie special \textit{initial conditions} for special \textit{dynamics}. What have we gained by this bargain? Whereas the initial conditions of inflation-less cosmology are plausibly viewed as tokens of extraordinary types, the \textit{dynamics} of inflationary models qualify as tokens of a less extraordinary type, given our theoretical background knowledge. 

Cynics will be inclined to attribute this verdict solely to our lack of a solid grasp of inflationary physics. Nescience about inflation’s particle-physical ``realisation” in particular is undoubtedly a major drawback. Nonetheless, the cynical aspersion belies the weight of the theoretical, particle-physical motivation for the inflationary framework. Albeit at present insufficiently constrained\footnote{\textcite[p.251, fn.6]{Smeenkthesis} rightly warns against \textit{overstating} the particle-physical motivation: ``(o)ne important disanalogy between [inflation and (speculative) GUT-scale physics] is that no fundamental principles guide inflationary model-building in the same sense that gauge invariance and renormalizability guide the unification program.”}, the empirically adequate models for inflationary dynamics---which include the \textit{simplest} ones (as \textcite{Linde:2014nna, Sabine2022} rightly underline)!---instantiate fairly generic types of dynamics---dynamics/potentials that our particle-physics background knowledge gives us reason to deem plausible.\footnote{Specific models can have further independent motivation. Starobinski inflation, for instance, can be viewed as the simplest, most conservative extension of GR within a Riemannian-geometric setting (see e.g.\ \textcite{Sotiriou:2008rp}). It has the additional advantage of being renormalisable.}

The spoils of the trade-off---of inflation’s ``explanation [of fine-tuned initial conditions] by subsumption under laws” \parencite[p.44]{Maudlin2007-MAUTMW}---are sizable at a thoroughly tangible, \textit{practical} level. Inflation’s answer to the fine-tuning calls for explanation unfolds its full force and relevance for the context of pursuit in virtue of its heuristic power. Whereas the reliance on initial conditions in the HBB-model remains invariably a contingent and barren input, the inflationary account thrives as a fertile research \textit{programme}: with its rich theoretical resources, it suggests new and powerful ideas and problem-solving strategies (cf.\ \textcite[p.69]{Worrall2002-WORHPT}; \textcite[Sect.5.2]{Seselja2014-EELEJI}. In this regard, ``(t)he mechanism for generating density perturbations is the most fruitful consequence of inflation” \parencite[p.206]{Smeenk2018-SMEIAT}---a topic to which we'll revert in \textbf{§7}. For the practising cosmologist, inflation’s great allure consists in its programmatic promise of advancing cosmological research, its \textit{vision forwards}; it ``immediately captivated” and ``engrossed'' \parencite[p.38]{steinhardt2011inflation} the majority of cosmologists.

\subsection{Explanatory Depth} 
Next, we'll examine the salient \textit{quality} that some of inflation's explanations possess---and that, conversely, the HBB-model lacks: explanatory depth. We'll unpack the notion in two senses in which explanations may be insensitive to changes in background conditions: one stresses robustness under causal interventions (\textbf{§4.2.1}) and the other stresses robustness under broader modal variations (\textbf{§4.2.2}). 

\subsubsection{Causal Depth}

The standard account of explanatory depth from \textcite{Hitchcock2003-HITEGP} is couched in a counter-factual theory of explanation (\textcite{Woodward2003-WOOEGP, Woodward2003-WOOMTH}): explanations are conceived as answers to ``what-if-things-had-been-different-questions” (w-questions). That is, ``explanation has to do with the exhibition of patterns of counterfactual dependence describing how the systems whose behavior we wish to explain would change under various conditions” \parencite[p.182]{Hitchcock2003-HITEGP}.

Woodward and Hitchcock's proposal for explanatory depth considers the range of such counterfactual variations under which the relation between the explanantia and the explananda remains invariant. An explanation’s \textit{causal}\footnote{Following Woodward \& Hitchcock in regarding counterfactual dependence under interventions as an explication of causal dependence, we’ll use ``causal depth” in demarcation from the modally broader form of explanatory depth in \textcite{Ylikoski2010-YLIDEP}, and discussed in \textbf{§4.2.2}.} depth is now defined as its degree of invariance under ``testing intervention”, i.e.\ the range of its generality, where the permissible counterfactuals are restricted to (physically) possible interventions or manipulations of the explanans variables. Interventions are physically (in principle) possible ways to change one variable, whilst holding other variables constant, thereby allowing us to determine its contributing effect on the explanandum. One may thus compare competing explanations with respect to their causal depth. Most important for our purposes is the range of their invariance (see
\textcite[p.184]{Hitchcock2003-HITEGP} for further details): is one explanation invariant under a larger set of interventions than the other? 

What makes explanations with great(er) depth attractive? Why deem it a dimension of explanatoriness---an explanatory value? As answers to w-questions, explanations cite reasons why the explanandum obtains---rather than some ``foil'' \parencite[p.204]{Ylikoski2010-YLIDEP}, or some member of a ``contrast class” \parencite[Ch.5]{VanFraassenBas1980-VANTSI}. Contrast classes generated by possible interventions involve physical possibilities with whose occurrence (or non-occurrence) one must reckon. We have sufficient certainty that these causal possibilities might easily have been actualised; in some idealised sense, they could even have been realised at will. Accordingly, we must consider them in most explanatory contexts: they typically should figure in the contrast class of satisfactory answers to a w-question; a good explanation ought to have something informative to say about the explanandum with respect to that contrast class. An explanation’s dimensions of invariance under counterfactual interventions thus express how satisfactorily it answers w-questions---against the welter of ``foils''. In particular, the broader an explanation’s range under possible interventions---the more causally robust the explanation---the more informative the explanation (see \textcite{Ylikoski2010-YLIDEP, Weslake2010-WESED} for details). Conversely, explanatory shallowness captures why some general claims are intuitively and commonly deprecated as explanatorily deficient: they fail to answer w-questions.

\subsubsection{Modal Robustness}

\textcite{Ylikoski2010-YLIDEP} generalise Woodward \& Hitchcock’s account of explanatory depth; we'll call their form of explanatory depth \textit{modal robustness}. Whereas the former emphasises invariance under causal changes, Ylikoski \& Kuorikoski consider invariance under modally broader counterfactual variations: the invariance demanded of explanatory generalisations extends to variations of the explanans parameters. Causal possibilities aren't the only ones that should figure in a satisfactory answer to a w-question---albeit for slightly different reasons (see below). 

The more \textit{sensitive} an explanation is to changes in its parameters, they suggest, the less ``robust'' or ``powerful'' it is. Less sensitivity implies that an explanation is fairly independent of those parameter-specific details; this is supposed to capture the intuition that ``good explanations make their explananda necessary or at least less contingent'' \parencite[p.208]{Ylikoski2010-YLIDEP}; necessary, that is, with respect to the relevant contrast class.

We'll be concerned with the range of values that the explanans parameters can take while still maintaining a viable explanatory relationship with the explanandum. A modally robust explanation is attractive for pragmatic reasons: it's a desideratum that, if satisfied, would facilitate the explanatory analysis. This becomes especially pressing under uncertainty regarding the actual values of the explanandum system's parameters---whenever we have insufficient knowledge to discard those parameter values. Accordingly, counterfactuals with respect to those parameters should figure in a good explanation's contrast-class (alongside causal possibilities). ``(S)ensitive explanations provide information that is unreliable in situations in which there are changes in factors that are not explicitly accounted for or when the case is extrapolated to unforeseen extremes'' \parencite[p.209]{Ylikoski2010-YLIDEP}. Conversely, more robust explanations can survive greater uncertainty with regard to the available epistemic access and empirical information we have. Such explanations are \textit{``structurally} stable'': the qualitative properties of the physical system don't sensitively depend on counterfactual variations in parameters within the explanans.

\subsection{Explanatory Vices and Virtues: The HBB-Model vs.\ Inflation} 
Let’s now apply these philosophical tools to inflation.\footnote{The results of our analysis are intended as an elaboration of what \textcite[p.28]{McCoy2015-MCCDIS} hints at, and a refinement of \textcite{Wolf:2022yvd} (see also \textcite{Azhar:2019qrj} for a quantitative account of explanatory depth with regard to inflation).} \textbf{§4.3.1} shows how the HBB model’s perceived flaws are naturally understood as explanatory shallowness. \textbf{§4.3.2} shows how inflation remedies those flaws: inflation is explanatorily deeper.

\subsubsection{The HBB-Model’s Vice: Explanatory Shallowness}
Recall the horizon and flatness problems and how a viable HBB-model must impose suitable initial conditions by hand. Now suppose---in analogy to systems in classical Statistical Mechanics---that changes in the universe’s initial conditions count as interventions (i.e.\ \textit{causal} possibilities, as conceptualised in \textbf{§4.2.1}).

Under these counterfactual variations, then, the HBB-model’s explanations for the universe’s flatness, uniformity, and primordial density perturbations break down cataclysmically. The explanations are sensitively dependent on very specific choices of initial conditions (see e.g.\ \textcite[Ch.5]{Mukhanov:2005sc}). The range of the invariance of the HBB-model’s explanations under interventions on the initial conditions is extremely limited. The HBB-model’s explanations are causally shallow: they fail to informatively answer salient w-questions. 

One may, however, sensibly contest this premise: according to \textcite[p.192]{Hitchcock2003-HITEGP} themselves (referring to the HBB-model's initial conditions), ``it is not clear that there are any well-defined testing interventions”. For initial conditions of the universe \textit{as a whole}, the idea of ``exogenous causal processes” \parencite[p.9]{Woodward2003-WOOEGP} via which to act on the universe doesn’t seem naturally applicable. The---for all we know!---uniqueness of the cosmos (cf.\ e.g.\ \textcite{Ellis2003TheUN}) exacerbates further doubts about causally effectible changes in initial conditions. Consequently, the HBB-model would \textit{trivially} lack causal-explanatory depth.

By the same token, if one construes changes in initial conditions instead as changes in parameters, the HBB-model’s explanations likewise come out as \textit{structurally} fragile---lacking depth qua modal robustness. Such variations in parameters differ in their ontological-modal import from causal variations: causal variations are plausibly viewed as possible, accidental changes of the target system---describing the \textit{same} system under different circumstances; modal variations, by contradistinction, lead to systems that, albeit similar, are nonetheless \textit{distinct} from the target system. One may therefore wonder why claims about such distinct systems should enter the ``contrast-class” figuring in explanations. The reasons are, as we'll elaborate in \textbf{§4.3.2}, more pragmatic; they facilitate model-building under uncertainty \parencite{tavakolXXXXfragility}. Structural stability is, even if perhaps not an inherent flaw, a desideratum---one that inflation, as we'll next show, delivers.

\subsubsection{Inflation's Virtue: Explanatory Depth}

Inflation provides deep explanations where the HBB-model dishes up shallow ones (for details,  recall \textbf{§\ref{inflation}}). 

\begin{itemize}

    \item Inflation's period of exponential expansion resolves the \textit{flatness problem}. It offers a robust mechanism that dynamically drives the universe's curvature towards zero, more or less regardless\footnote{The full extent of initial conditions that allow for inflation to begin is still not conclusively settled (see e.g.\ \textcite{Brandenberger:2016uzh, East:2015ggf}). While inflation cannot begin in \textit{any} completely arbitrary initial state, inflation can begin for a wide range of initial conditions, even those that are surprisingly inhomogeneous.} of the universe's initial curvature and mass-energy density.

    \item Inflation's period of exponential expansion resolves the \textit{horizon problem}. The homogeneous regions in the sky’s CMB that prima facie appear too remote from each other for any past causal contact, on the inflationary picture, grew out of a smaller, causally connected patch---before inflation started. Inflation then rapidly inflates away inhomogeneities.

    \item By the same token, inflation offers a deep \textit{explanatory link between cosmic structure and the density perturbations} generated by the inflaton. Inflation causally explains the origin of these density perturbations and their \textit{superhorizon correlations}. Their statistical properties---the approximate scale-invariance of the observed power spectra in particular--- must be accepted as brute facts within the HBB-model. By contrast, inflation generically predicts it---a prediction that doesn’t sensitively hinge on the initial matter density or distribution.

\end{itemize}
%In contrast with the explanations offered by the HBB model, which were found to be shallow, these inflationary explanations can be understood to have modal depth. 
As before, one can sensibly contest the idea that there are any well-defined causal interventions on the universe as a whole. Yet, we can still consider modal variations of parameters such as the mass-energy density and distribution. The inflationary explanations do not sensitively depend on these parameters and are thus structurally stable: they remains intact under a wide range of counterfactual variations of the parameters.

Inflation supplies deep explanations also beyond these remedies of the HBB-model’s explanatory blemishes. One in particular stands out: the historically influential \textit{monopole problem} \parencite[Ch.9\&10]{guth1997inflationary}.\footnote{Albeit routinely mentioned in standard presentations recently, its popularity has somewhat waned. In part, this is owed to the growing disillusionment over the absence of evidence for physics beyond the standard model of particle physics, for GUTs in particular (see e.g.\ \textcite{Hossenfelder:2018jew}); in part, it reflects a shift in emphasis towards more compelling---in particular, empirical—arguments for (or ``earmarks'' of inflation (see \textbf{§7}). Given its traditionally accorded importance, the monopole problem nonetheless deserves discussion in the context of pursuit, especially in our rational reconstruction of inflation’s quick ascension to the dominant paradigm within cosmology.} It’s best described as an ``external'' problem (see \textcite{Penrose:1988mg}; \textcite[fn.4]{McCoy2015-MCCDIS}), rather than an internal one: when combined, the HBB-model and GUT physics clash. GUTs---setting aside their iffy status---generically predict certain relic particles (``topological defects”), such as magnetic monopoles or cosmic strings. To-date, though, they have eluded detection. This non-observation, then, must be attributed either to (i) a statistical fluke, or (ii) to having considered the wrong GUTs (presuming that GUTs exist). The former option is evidently explanatorily trite: qua probabilistic quantum theories, GUTs never rule out such coincidences. The second option, (ii) is modally fragile in a more interesting sense. Most garden-variety GUTs predict copious amounts of topological defects. Strategy (ii) sensitively depends on non-standard assumptions about more fundamental (viz. GUT-scale) physics. 

Inflation elegantly solves the monopole problem by diluting any possible relic particles from GUTs, reducing their density below detectability. For the most part, this doesn't depend on any details about GUTs. The low-density of relic particles is modally robust: it remains intact under broad\footnote{NB: Counter-nomic variations are, of course, modally even more remote possibilities than the counterfactuals considered in \textbf{§4.2.2}.} modal---viz.\ counter-\textit{nomic} with respect to laws at the GUT-scale)---variations (an instance of ``autonomy” in the parlance of \textcite{Wolf:2022yvd}).

We close this section with a discussion of the value of inflation’s modal robustness---the explanation’s invariance under counterfactual variation of initial conditions or parameters of the universe. 
%Owing to its direct link with relevant contrast-classes, causal depth is squarely related to an explanation’s \textit{intrinsic} explanatory quality. By contrast, modal robustness---the explanation’s invariance under counterfactual variation of initial conditions or parameters of the universe, or even counternomic variation---is different: 
For all we know, we inhabit a unique universe, characterised via certain contingent parameter values (or laws). Why care about non-actual values (or laws), corresponding to different universes? 

The advantages of a structurally stable---and, conversely, the \textit{dis}advantages of structurally \textit{un}stable/fragile---explanations are, as \textcite[p.31]{McCoy2015-MCCDIS} underscores, \textit{pragmatic}: ``(a)s a matter of risk reduction in theory construction, theorists would much prefer to hedge their bets on a theory with greater explanatory resources”. Of course, ``[...] real observations always involve some degree of inaccuracy [...] So long as the model accurately representing the real world is nearby (`a perturbation away'), then we can expect that predictions based on using the `inaccurate' model will be approximately correct” \parencite[p.84]{McCoy2020-MCCSIC}. By wedding cosmological theorising to highly specific choices of initial conditions and/or parameters characterising our universe, one gratuitously sticks one’s neck out. Furthermore, these advantages bear directly on knowledge that we presently have and our potential epistemic access to further knowledge. That is, there are good reasons to think that further empirical access to such parameters will be very limited. The nature of cosmological research is uniquely constrained: we can never peer beyond (or rather before) the surface of last scattering observed in the CMB. Consequently, we'll almost certainly never have direct access to these parameters at the beginning, making such highly specific choices even more precarious. 

%By avoiding the need to make such specific and finely-tuned choices, inflation offers a far more secure explanation for the observed state of the universe.

Prudence counsels caution in theory-pursuit instead: we ought to prefer research strategies that minimise such hazardous commitments. Structural stability achieves just that: it permits us, as we ineluctably must, to work with approximations, uncertainty, error-contaminated data models, and idealisations (cf.\ especially \textcite{Norton2012-NORAAI})---without having to fret over potentially momentous and epistemically uncontrollable consequences.\footnote{\textcite[Sect.5]{1992GReGr..24..835C} have drawn  attention to the utility, if not practical indispensability, of \textit{presuming} stability in order to gain computational control over especially cosmological perturbations.}

\section{\textit{Promise 2:} Unification}\label{unification}

\noindent In an otherwise sceptical article, \textcite[p.262]{unruh1997} admits: ``[…] I am attracted, as are most people, by the beguiling promise of the field. With one theory one can explain so much more than one would ever have expected to explain.” Here, we’ll unravel this sentiment---inflation’s astonishing capacity for unification. Unification is, we maintain (following especially \textcite{Kao2019-KAOUBJ-2}), a highly coveted feature in a theory; its display justifies a theory's pursuit. This section will argue that inflation indeed promises unification (or, more precisely, various facets of unification), and hence accrues further pursuit-worthiness---especially in the absence of alternatives holding out an equal promise. 

\subsection{Forms of Unification}
Let’s begin with distinguishing several different aspects, or forms, of unification, salient to the case of inflation.\footnote{We borrow the taxonomy (with minor adjustments) from \textcite{Falkenburg2012-FALPUO} (see also \textcite[Ch.IX.E]{Bartelborth1996-BARSE-2} for a similar one from a more general perspective).}

\begin{itemize}
    \item \textit{Methodological} unification involves the transfer of methods from one theory to another.\footnote{See \textcite{Nyrup2020} for a philosophical analysis of the pay-offs especially for the context of pursuit.} 
Such methods can be experimental or those of data analysis. But they may naturally also include computational or mathematical methods (e.g.\ Renormalisation Group Methods), analogical or heuristic reasoning with concomitant modelling techniques (e.g.\ Wiener processes from statistical mechanics in stock exchange models).
\item \textit{Phenomenological} unification occurs when disparate phenomena are subsumed under an overarching principle. The antecedently disconnected phenomena thus are ordered via some parameters, or more abstract structures. An example is the ``all particle spectrum” in astroparticle physics; it arranges all cosmic rays, according to their (experimentally determined) energy fluxes, irrespective of their types or likely origins. 
\item \textit{Conceptual} unification\footnote{Or, as \textcite{Morrison:2013hjc} puts it, ``synthetic unity''.} is achieved when a conceptual framework is forthcoming within which one can describe and model the phenomena. Einstein-Maxwell theory is a case in point: it allows a description of electromagnetism within the framework of general-relativistic spacetimes. Likewise, arguably the standard model of elementary particle physics provides a conceptual unification of the electroweak and the strong force only in the sense of a description in a common language and mathematical structure (\textcite{Maudlin1996-MAUOTU-2}; \textcite[Sect.3]{Morrison:2013hjc}).
\item \textit{Explanatory} unification seeks a unified account of the phenomena: it’s their explanation that unifies or systematises them. While the notions of unification and unificatory power have received much attention in the philosophical literature (e.g.\ \textcite{Kitcher1989-KITEUA, Friedman1974-FRIEAS, Morrison2000-MORUST-2}), we’ll work with the following rough characterisation in terms of coherence \parencite{Bartelborth1999-BARCAE, Seselja2014-EELEJI,BonJour1985-BONTSO-4}. It can be both internal and external. 

Internal coherence means that the corpus of explanantia has a tight inner structure: its various elements are well-connected, and form an organic system. An example is given by a common causal mechanism responsible for producing various phenomena. Three principal dimensions constitute a theory’s internal coherence (see \textcite{Bartelborth1999-BARCAE, Bartelborth2002-BAREU-2} for further details). One lies in its capacity to systematise a wide array of phenomena, the theory’s scope: the more applications a theory has, the greater its unificatory power. A second dimension is given by the theory’s empirical content: the more specific a theory is---the more situations it forbids---the more informative, and hence greater, its unification. A third dimension pertains to its organic, unified internal structure, the reliance on as few independent postulates, laws, etc. (as opposed to merely via conjunction): a theory scores high in this regard, if it uses only few principles or patterns/regularities or if they naturally hang together, with multiple and strong links amongst the various elements of the theory. 

External coherence means that the corpus of explanantia has tight links with \textit{other} theories. At a minimum, they ought to be mutually consistent; ideally the explanantia mesh with our background knowledge, leading, for instance, to a fully satisfactory explanation through other parts of physics.

\end{itemize}

\subsection{Inflation and Unification}

Cosmic inflation exhibits unificatory power along all four foregoing aspects.

It achieves \textit{methodological unification} by combining ideas and techniques from both cosmology and quantum field theory and particle-physics (see \textcite[Sect.III]{Baumann2009}). Especially relevant in this regard is the reliance, in inflationary cosmology, on quantum field theory on curved spacetime, an extension of the standard quantum field framework to situations with non-negligible (general-relativistic) gravity. Furthermore, inflation borrows its key concepts directly from well-entrenched particle physics: the idea of phase transitions via spontaneous symmetry breaking and a false vacuum state (see e.g.\ \textcite[Ch.8\&10]{guth1997inflationary}). But also more computational techniques are important: perturbation theory both within GR, and standard quantum field theory (on a Minkowskian background), applied to curved spacetimes, are (triumphantly!) exploited in determining the primordial density perturbation spectrum---an application that feeds into three other dimensions of unification. 

Inflation brings about \textit{phenomenological unification} by linking various previously unrelated phenomena. The classical perceived flaws the HBB-model (the fine-tuning issues, alongside the monopole problem, recall \textbf{§4}) are an immediate case in point: they concern sundry, per se unconnected, and individually contingent features of initial conditions. According to inflation, they, astoundingly, turn out to be connected. Inflation in fact provides an explanation of a common origin. 

Secondly, inflation connects staggeringly different scales. ``If inflation is right, the intricate pattern of galaxies and clusters of galaxies may be the product of quantum processes in the early universe. The same Heisenberg uncertainty principle that governs the behavior of electrons and quarks may also be responsible for Andromeda and the Great Wall'' \parencite[p.216]{guth1997inflationary}! 

Thirdly, and likewise by tracing them back to a common origin, inflation links phenomena as disparate as information in the CMB---the faint afterglow of the plasma of photons from 380,000 years after the Big Bang---the present-day  distribution of cosmic large-scale structures, such as galaxy clusters, and patterns in primordial gravitational radiation, ripples in spacetime itself created during the first second after the Big Bang (see also \textbf{§7.2}).
%%This is not exactly correct. We always knew that perturbations in the CMB were linked to large scale structure. That is why finding their amplitude was so important and why this evidence was hinting at the existence of dark matter, quite independent on inflation. REPLY: AGREED---THE VERB WAS TOO STRONG: IT MISLEADINGLY SUGGESTED NOVELTY (WHICH IS HOWEVER NOT THE SUBJECT OF THIS SECTION). NONETHELESS, THE UNIFICATION OF SUCH DIVERSE PHENOMENA---WHETHER EXPECTED OR NOT---IS IMPORTANT: BEAUTIFULLY ILLUSTRATES HOW SUCH BIZARRELY DIFFERENT PHENOMENA ARE LINKED! I FIND THIS PHENOMENOLOGICAL UNIFICATION TREMENDOUSLY IMPRESSIVE.  

Inflation \textit{conceptually unifies} the various phenomena by providing the framework within which one can describe---and, of course, explain---them: the framework comprises on the one hand, classical GR, with ``exotic matter”, and on the other hand the framework of quantum field theory on curved spacetimes.\footnote{Philosophically, the application of quantum-mechanical ideas to the universe as a whole is, of course, particularly intriguing in light of the measurement problem (see e.g.\ \textcite{Wallace}) and the role and significance of decoherence. We’ll set these questions aside here (see however e.g.\ \textcite{Martin:2018zbe}).} What is distinctive of inflation’s conceptual unification is that it morphs contingent posits---specially chosen initial conditions---into fairly generic consequences of new laws. In \textbf{§4.1}, we saw how thereby inflation generates its puissant ``positive heuristic''.

Finally, inflation’s \textit{explanatory unification} consists in providing a \textit{common mechanism} for the phenomena. \textcite[p.459]{Janssen2002-JANCSE} has forcefully stressed the role of such a type of unification as a general methodological pattern, successful and pervasive throughout the history of science: one that traces  ``(traces) striking coincidences back to common origins”.\footnote{Note that Janssen expressly deems the unification accomplished through a common origins inference \textit{evidence}. In line with our overarching agenda, we only need a weaker premise:  unification is such an enticing promise---through enhancing understanding, it realises a goal of science---that theories with unifying capacities merit \textit{pursuit}. Recently (as reported by \textcite[p.3266, fn.3]{Kao2019-KAOUBJ-2}) Janssen has shifted his view, now stressing the postulate of a common origin of several problems as an ``indicator that an idea is worth pursuing''.}  As \textcite[fn.58]{Smeenk2005-SMEQVE} observes, inflation neatly dovetails such unificatory reasoning. Inflation, as the ``common origins inference'', displays the internal and external coherence that make its explanatory unification powerful along the dimensions specified above.
\begin{itemize}
\item As per its phenomenological unification, it has a plethora of applications, on diverse scales and involving highly variegated phenomena.
\item Making several specific predictions (more on this in \textbf{§7}), inflation has substantial empirical content: it asserts \textit{informative} connections amongst the phenomena within its domain. 
\item Inflation’s internal structure emanates from its central simple idea: an exponential expansion of the universe. Together with its particle-physical motivation, inflation can be said to form an organic whole---despite lingering uncertainties regarding its specific particle-physical realisation, and other open problems.   
\item It has---or rather: forges---tight links with other theories: particle-physics\footnote{This isn’t to deny, as \textcite[p.218]{Smeenkthesis} stresses, that those links aren’t without problems. But focusing on inflation’s promise-based pursuit-worthiness, we’ll not discuss them further here.} (both well-established as well as more speculative forms, such as string theory or supersymmetry), GR and relativistic cosmology, astrophysics. Especially interesting in the context of pursuit are two benefits of such links. One concerns the cross-fertilisation of different areas (as per methodological unification). It kits researchers with a versatile and ample arsenal of tools and ideas to work with. The other concerns the prospect of new tests of either applications of established theories in new domains, or of so-far unconfirmed theories beyond the reach of direct tests (see e.g.\ \textcite{Smeenk2005-SMEQVE} for details).\footnote{In this vein, \textcite[p.120]{EllisUzan} underscore the significance of linking cosmology and nuclear physics (culminating in the triumphant incorporation of primordial nucleosynthesis into the standard model of cosmology): ``(u)ntil the 1960s, most physicists thought cosmology was just philosophy, hardly worth taking seriously. That changed first when atomic physics became relevant to the universe through Gamow's realization that a hot Big Bang early phase must have occurred [...]. It is this that made cosmology a respectable physical science [...].'' As they also stress, the significance was \textit{mutual}: cosmology opened up new testing grounds, and constraining data for nuclear physics (illustrated, for instance, by the restriction of the number of neutrino families).}
\end{itemize}

\section{\textit{Promise 3:} Understanding}\label{understanding}

\noindent Does cosmic inflation furnish a degree of understanding the relevant cosmological data, else not available? In this section, we’ll explore the idea that much of inflation's enticement consists in a positive answer. Understanding here is conceived of distinct from knowledge (or acceptable beliefs), but still ``a generally acknowledged aim of science” \parencite[p.165]{deRegt2005-REGACA-2}. Insofar as inflation promises to promote this aim, it merits pursuit---especially when it promises this aim more so than alternatives. 

It lies outside of our paper’s ambit to defend any substantive position within the burgeoning discourse on understanding. Instead, we’ll cull from the literature (see e.g.\ \textcite{Lipton2009-LIPUWE, Baumberger2017-BAUWIU, Grimm2011-GRIU, Hannon2021-HANRWI-2}) those commonly mentioned aspects that, we think, make understanding especially salient for cosmic inflation.  

Let’s begin by noting (without discussion, see e.g.\ \textcite[Sect.2]{Gordon2017-AUTUIE} for a review) that the connection between understanding and truth (or compliance with acceptability criteria) tends to be seen as weaker than that required for knowledge. Moreover, in contrast to knowledge, understanding is usually viewed as compatible with epistemic luck \parencite{Gordon2017-AUTUIE,Baumberger2017-BAUWIU}. In light of the tenuous evidential situation of cosmic inflation until (at least) long after its endorsement through the scientific community, this renders understanding an auspicious candidate for inflation’s virtues.

\subsection{Forms of Understanding}
How to construe ``understanding”? What does it amount to? Which features of an explanation, or a style of reasoning, confer understanding? Frequently found in the literature are four main (mutually not exclusive) suggestions.
\begin{enumerate}
    \item \textit{Coherence and unification}:  Understanding is constituted by coherence-making relationships in a body of information. We understand when we ``grasp” how pieces of information hang together. We achieve understanding through unification (cf.\ \textcite{Bartelborth1999-BARCAE,Kitcher1989-KITEUA}). For our purposes, a rough gloss of unification shall suffice (recall \textbf{§5}): the more (otherwise contingent) facts it subsumes under as few law-like regularities as possible, the more unified the account of those facts is.
    \item \textit{Causal mechanism}: ``How things hang together'' can be construed also slightly differently---by emphasising causal mechanisms. On this reading, understanding is constituted by revealing a phenomenon’s relevant causal-mechanical processes. Without embroiling ourselves in the tangles of controversial metaphysics, again a rough gloss shall suffice (see \textcite{Craver} for details): a causal mechanism exhibits the fundamental-physical dynamical interactions of matter that produce the thing-to-be-understood.
   \item \textit{Ability to make counterfactual evaluations}: On this proposal, one construes the grasp of how things hang together in yet a different way. It stresses the ability to answer w-questions (\textbf{§4.2})), ``the ability to anticipate the sort of change that would result if the factors cited as explanatory were different in various ways” \parencite[p.12]{Baumberger2017-BAUWIU}. Thus, to understand a phenomenon (via a theory), is constituted by our apprehension of the fundamental dependency relations, characteristic of the phenomenon-to-be-understood; we understand when we can competently evaluate counterfactuals situations---when, in other words, the theory enables us to give an answer-to-a-why question, relative to a contrast class.
    \item \textit{Intelligibility}: Lastly, one may construe understanding in terms of intelligibility, a good ``grip of its gist”. A theory $T$ confers understanding when it enables its competent users to ``recognise qualitatively characteristic consequences of $T$ without performing exact calculation'' \parencite[p.151]{deRegt2005-REGACA-2}. This kind of understanding helps us to straightforwardly \textit{intuit} the consequences of the theory.

\end{enumerate}

\subsection{Inflation and Understanding}
   
Drawing on the results of earlier sections, it’s immediate to glean how cosmic inflation indeed enhances our understanding thus understood. 

\begin{itemize}
    \item Several researchers have extolled the capacities of inflation for unification as a salient virtue. We can unpack this along two dimensions. The first is summed up by \textcite[p.19]{EllisUzan}: ``[…] one of the major driving forces of physics for the past several hundred years has been to unify apparently distinct physical phenomena by giving a single explanation. […] What we really want is a proposed mechanism that is not just used to explain one phenomenon (inflation) but also \textit{several phenomena in different contexts}.” We discussed this in detail in \textbf{§5}.

    A second dimension of unification may, following \textcite[Sect.2]{Mikowski2016-MIKFAC}, be called ``integration''. It generalises the specific kind of unification---through a common origins inference---to a broader bringing together of general-relativistic cosmology and other areas of physics. \textcite[p.20]{EllisUzan} again hint at the idea: ``(t)he dream of \textit{linking} particle physics to the very early universe was a high hope when inflation was proposed”.  

    \item Throughout the literature on inflation, the desirability of a causal-mechanical account is explicitly affirmed. \textcite[p.428]{Hobson:2006GL})'s introduction to inflation is characteristic: ``[...] we saw that standard cosmological models suffer, in particular, from the flatness problem and the horizon problem. [...] one would hope to explain this phenomenon with an underlying physical mechanism.”\footnote{For a philosopher’s explicit endorsement of the desirability of a causal-mechanical account along the lines we sketched, see \textcite[Ch.1]{Maudlin2007-MAUTMW}.} 

    \item With respect to understanding as the ability to make counterfactual evaluations, cosmic inflation scores superbly.\footnote{We aren't aware of passages in the physics literature that clearly state a link between understanding, thus construed, and the---well-documented---ubiquitous appreciation of inflation's modal robustness (see \textcite[p.144]{Peebles2022} for coming close to being an exception). Still, we believe that the model in question plausibly qualifies as a  rational, even if not always articulated, reconstruction of that appreciation. In other domains of physics, as pointed out in the literature on understanding, the model reflects widespread intuitions.} Thanks to its explanatory depth (\textbf{§\ref{depth}}) evaluating counterfactuals---understood standardly as conditionals in which one varies initial conditions, whilst keeping the physical laws intact---is particularly easy: generically, counterfactuals won't be affected!

    \item  Intelligibility in the sense of intuitive grasp---responsible for eliciting a moment of `aha!'---captures a virtue routinely extolled in discussions of inflation. \textcite[p.35, our emphases]{Guth:2002}, for instance, writes: ``(w)ithout inflation, general-relativistic cosmology faces the ``\textit{difficulty of understanding} the large-scale homogeneity of the universe''. Moreover, ``(t)he  inflationary model also provides a \textit{simple} resolution of the flatness problem [...]''.  Its standard, intuitive illustration is found in almost every popular exposition of  cosmic inflation: the surface of an extended sphere becomes geometrically flatter as it gets larger.

    Furthermore, two additional arguments for inflation that \textcite[Sect.1.3]{Guth:2004tw} lists are naturally interpreted in terms of understanding as affording a qualitative, intuitive grasp. First, the universe is incredible large and contains vast quantities of matter. Inflation, through inducing exponential expansion, quite simply and intuitively explains this size. Secondly, the inflation-less HBB-model has to take the observed cosmic expansion as a further postulate of the initial conditions (what \textcite[p.248]{Hobson:2006GL} call the ``expansion problem''). ``It proposes no answer at all to the question of what banged, how it banged, or what caused it to bang'' \parencite[p.236]{guth1997inflationary}. Cosmic inflation offers ``just the kind of force needed to propel the universe into a pattern of motion in which each pair of particles is moving apart with a velocity proportional to their separation” \parencite[p.6]{Guth:2004tw}. 

\end{itemize}

\section{\textit{Reality Checks:} Prospects of Testing}\label{novelty}    

\noindent So-far, we have inspected the promises of inflation, evinced by its salient, primarily \text{super}-empirical theory virtues. This section will complementarily analyse a more empirical one: inflation makes certain generic predictions (\textbf{§\ref{predictions}}). 
In the case for its pursuit-worthiness, this fulfills two functions, since researchers, \textgreek{ὅς ἐπὶ τὸ πολύ}, \textit{hope} that a theory they are pursuing will receive empirical confirmation, and accordingly eventually earn acceptance. First, predictive power allows physicists to ascertain not too far down the road whether inflation is on the right track. Predictions---either already performed or at least in the offing---serve as preliminary reality or plausibility checks: they act as pro tem touchstones for adjudicating whether inflation's promises warrant a modicum of trust. Secondly, such plausibility checks seem especially compelling for \textit{novel} predictions---as inflation indeed makes(\textbf{§\ref{novel}}). These predictions are novel in (at least) two important senses. This demarcates inflation from alternative proposals: they share with inflation several of the virtues discussed in the preceding sections; yet they lack inflation's predictive novelty. Thereby inflation gets an extra edge in terms of pursuit-worthiness.

\subsection{Predictions}\label{predictions}
It verges on a philosophical platitude that scientific theories require  empirical tests, usually in the form of predictions. One of the rationales behind that demand is that the passing of such tests is supposed to redound to a theory's credibility; the theory thereby receives some special kind of confirmation. While this paper focuses on the pursuit-worthiness of inflation---rather than the evidential credentials for its acceptance---it seems reasonable (though not uncontroversially so, cf.\ \textcite{Shaw2022-SHAOTV-2}) to enjoin that a pursuit-worthy theory \textit{be able} to pass a certain evidential threshold in the near future (or better still, to pass it already): for a theory to be pursuit-worthy one ought to be able to figure out whether in due course it \textit{might} qualify as acceptable. Ideally, in other words, a pursuit-worthy theory should make at least some rough predictions that can be checked; the prospects of tests crucially contribute to a theory's pursuit-worthiness.

Inflation makes a handful of generic predictions \parencite{Linde:2014nna, Guth:2013sya}; some of them can be---and in fact \textit{have been}---tested. We encountered two\footnote{Another prediction has also been alluded to in passing: inflation predicts the existence of tensor perturbations in the form of primordial gravitational waves, as well as particular statistical features for this tensor power spectrum (see e.g.\ \textcite[Sect.III]{Baumann2009} or \textcite{Guzzetti:2016mkm}). Although the predicted amplitude of such signals is model-dependent, the prediction of a nearly scale-invariant power spectrum for these tensor perturbations is a similarly robust result within the general inflation paradigm. These signals haven't yet been detected, but the null results only rule out a small segment of inflationary models \parencite{Planck:2018jri}. The confirmation of this prediction is a major target for experimental cosmology.} of those predictions already:    
\begin{enumerate}
    \item The near-perfect geometric flatness of the universe, with curvature deviating from flatness at a quantifiable (micro-percentage) level.
    \item The near, but not exact, scale invariance of the density perturbation power spectrum. 
\end{enumerate}
We’ll elaborate on them below. They illustrate our main point here: what makes them so interesting for inflation’s pursuit-worthiness is their novelty.

\subsection{Predictive Novelty}\label{novel}

Alternatives to cosmic inflation exist, such as bouncing cosmologies and string gas cosmology (see e.g.\ \textcite{Brandenberger:2011et}; \textcite{Brandenberger:2016vhg}). Most of them were devised much later, though. Their advocates may rightly boast that they share with inflation several of same virtues as inflation. The virtues of many bouncing cosmological models closely resemble those of inflation with respect to depth (but with some important differences, see \textcite{Wolf:2022yvd}), as well as unification and understanding. But also alternatives that depart more radically from the standard inflationary framework, such as string gas cosmology, arguably exhibit those virtues. Following our reasoning in \textbf{§\ref{pursuit}-§\ref{understanding}}, those alternative proposals consequently seem to hold similar promise. Should they be pursued with \textit{equal} vigor?\footnote{Obviously, the question of inflation's \textit{historical} pursuit-worthiness prior to the development of those alternatives becomes moot.} We don't think so: vis-à-vis those alternatives, inflation stands out in making (having made) \textit{novel} predictions. 

A theory's predictive novelty, its ability to make novel predictions, is widely prized as arguably ``the single most important theoretical virtue'' \parencite[p.69]{Schindler2018-SCHTVI-5}. Typically, it's invoked as support for a theory---i.e.\ in the context of confirmation (\textbf{§3.2}). For our purposes, a weaker and intuitively plausible claim will do (cf.\ \textcite{Douglas2013-DOUSOT}). Consider a situation where a theory has non-trivial predictive novelty, but falls short of the customary standards for confirmation. Under these circumstances, we'll presume, a theory's predictive novelty still signals greater promise than one lacking it. Predictive novelty provides epistemic assurance, boosting the theory's pursuit-worthiness: it vouchsafes a minimal plausibility that one may reasonably require of a theory expected, or hoped, to pass at some point the usual evidential standards of acceptable theories. Applying this reasoning to inflation, we maintain that thanks to its predictive novelty, inflation should be preferred over the inflation-less HBB-model or competitor theories (both of which lack predictive novelty). 

Various proposals exist for cashing out predictive novelty (see e.g.\ \textcite{Musgrave1974-MUSLVH}; \textcite{Carrier1988-CARONF-3}; \textcite[Ch.3]{Schindler2018-SCHACC-4}).\footnote{They include \textit{temporal} novelty, \textit{heuristic/use}-novelty, \textit{problem} novelty, \textit{theoretical} novelty, and \textit{non-adhoc} novelty. We believe that inflation scores highly on \textit{all} of these accounts. A companion paper will explore the subject in greater detail.} For the present purposes, two are particularly germane. 
Without adopting a partisan attitude towards them, we'll briefly consider how inflation fares \textit{vis-à-vis} \textit{temporal} (§\ref{temporal}) and \textit{problem} novelty (§\ref{problem}). Inflation, we'll argue, is predictively novel in both senses.

\subsubsection{\textit{Temporal} Novelty}\label{temporal}

Traditionally, as championed by e.g.\ \textcite[p.36]{Popper1963-POPCAR} or \textcite[p.5]{Lakatos1978-LAKTMO}, novelty has been construed in temporal terms: a prediction only counts as novel, on this view, if the forecast fact or phenomenon hasn't been \textit{known beforehand} (or has even been expected \textit{not} to occur). 

When inflation was developed, the value of the universe's spatial curvature was only weakly constrained.  \textcite[p.347]{Guth:1980zm} noted at the time that ``one can safely assume'' that $0.01 < \Omega <10$. Until the early 2000s, an open universe still remained very much an empirical possibility;  many prominent cosmologists  in fact favoured it \parencite{peebles1986mean, coles1997case}. Inflation not merely explains why the universe is somewhat flat to within a couple orders of magnitude (i.e.\ solve the flatness problem as it was understood in the 1980s); it gives a specific prediction that the present value of $\Omega$ in our universe is $\Omega \simeq 1$.\footnote{Some physicists (e.g. \textcite{Bucher:1994gb}) and philosophers (e.g. \textcite{Earman1999-EARACL}) have questioned whether flatness is genuinely a robust prediction of inflation: it's possible to construct open models of inflation, which allow for non-flat (especially open) universes. Nonetheless, a flat universe can be said to be a generic prediction of inflation, whereas an open universe can only be \textit{accommodated}. According to this distinction (which draws on \textcite{Worrall1985, Worrall2014-WORPAA}), predictions are achievements of a theory to account for facts in a manner that flows from the theory's natural resources;  accommodations, by contrast, occurs when data itself is used to determine parameters (or free functions) in the theory. In this sense, inflation can accommodate an open universe by using the observed curvature in an open universe as input data to reverse engineer/fix appropriate field values during the inflationary stage (e.g.\ see \textcite{Bucher:1994gb}). By contradistinction, the basic structure and dynamics of inflation quite \textit{generically} imply a flat universe; such a particular data point is \textit{not} harnessed for specifying any parameters/free functions within the theory. A follow-up paper will provide a more comprehensive analysis.} This observation was first partially confirmed when the Boomerang mission detected a peak in the angular power spectrum of the CMB, giving $ 0.88 < \Omega < 1.12$ \parencite{Boomerang:2000efg}. Subsequent CMB experiments WMAP and PLANCK further refined this value, to currrently $1-\Omega = 0.0007 \pm 0.0019$ \parencite{Planck:2018vyg,WMAP}. 
Inflation's prediction of a geometrically flat universe anticipated these measurements by approximately two decades \parencite{Guth:1980zm}.

Proceeding to cosmic structure, ``the most remarkable feature of inflation, widely recognized shortly after Guth’s paper, was its ability to generate a nearly scale-invariant spectrum of density perturbations with correlations on length scales larger than the Hubble radius'' \parencite[p.216]{Smeenk:2017uof}. Inflation gave a clear prediction that the power spectrum of these density perturbations should be \textit{nearly}, but \textit{not exactly}, scale-invariant (as well as adiabatic and Gaussian, recall \textbf{§2} for details) before the primordial density perturbations themselves had even been detected.\footnote{The first measurements came from the COBE satellite (Smoot et al., 1992).} The first confirmation of inflation's predictions for the power spectrum of the density fluctuations came from the WMAP satellite; it ruled out a perfectly scale-invariant spectrum \parencite{WMAP}. PLANCK further corroborated the finding by measuring $n_s = 0.9649\pm 0.0042$---in excellent agreement with what one would expect if an inflationary epoch sourced these perturbations \parencite{Planck:2018vyg}. As with flatness, inflation's successful prediction preceded the observations by decades (\textcite{Mukhanov:1981xt, bardeen1983, guthpi1982, Hawking:1982cz}, see \textcite{Smeenk2018-SMEIAT} for historical details).

\subsubsection{\textit{Problem} Novelty}\label{problem}

\textcite[p.2]{Gardner1982-GARPNF} delineates a distinct nuance of predictive use-novelty. It pivots on what he dubs ``\textit{problem}-novelty”: problem-novel phenomena ``don’t belong to the problem-situation which governed the construction of the hypothesis”. That is, problem-novel phenomena don’t belong to the \textit{class of problems} that the theory’s inventor considered their theory responsible to solve. For a theory's prediction to exhibit this type of novelty, the theory must not be ``specifically designed to deal with the facts'' or ``cleverly engineered'' to reproduce them \parencite[pp.102]{Zahar1973-ZAHWDE-2}. 

A glance at the earliest papers on inflation verifies that inflation wasn't specifically designed to predict the geometric flatness of the universe, or to source cosmic structure. Rather, both \textcite{Guth:1980zm, Starobinsky:1980te} used ideas from particle physics to investigate how cosmological models might behave at the higher energy-scales in the early universe. While Guth initially investigated phase transitions in GUT models and later had a ``spectacular realization'' regarding the resolution of fine-tuning issues \parencite[p.179]{guth1997inflationary}, Starobinsky developed models of inflation from a different perspective. He argued that we should expect quantum-mechanical corrections to GR in the high-curvature, high-energy regimes of the early universe. Consequently, he pursued the cosmological consequences that could result from these corrections and similarly found that we should expect a de Sitter-type exponential expansion in the early universe. Unlike Guth, Starobinsky in fact didn't notice any connection between this behaviour and the resolution of the HBB-model's fine-tuning problems. Regarding cosmic structure formation, as we already commented on, it was only \textit{after} inflation's invention that it was realised that the theory provided such one. As \textcite[p.336]{Baumann:2022mni} emphasises, ``[...] the theory was not engineered to produce these fluctuations, but that their origin is instead a natural consequence of treating inflation quantum mechanically.'' 

Both of these predictions are problem-novel (see also \textcite[Ch.7.3]{Smeenkthesis}): the theorists who originally developed inflation weren't striving for a theory that resolves the fine-tuning problems or that accounts for cosmic structure formation; rather they \textit{later} noticed that applying well-motivated ideas in high-energy particle physics to cosmology naturally offered compelling answers to those questions. 

In sum: Irrespective of whether one construes predictive novelty in temporal or problem-novel terms, inflation has a significant leg up on its competitors. More recent alternatives obviously cannot compete with inflation in terms of temporal novelty. Furthermore, they were developed in a context in which the problems that inflation unexpectedly solved were more fully integrated into the problem situation; hence they can't be said to be problem-novel either.

\section{Conclusion}\label{conclusion}

\noindent \textcite[p.118]{Guth:2013sya} characterise inflation as ``a self-consistent framework with which we may explain several empirical features of our observed universe to very good precision, while continuing to pursue long-standing questions about the dynamics and evolution of our universe at energy scales that have, to date, eluded direct observation.'' In light of our analysis, we can concur with this characterisation: vis-à-vis its simultaneous display of multiple, salient theory virtues, it comes as no surprise that inflation is and remains actively pursued as the dominant framework for modeling the early universe. The reasoning in this paper, we hope, has succeeded in explicating what, as \textcite[p.6]{rees1997} puts it, makes ``the inflation concept [...] the most important idea of the last 20 years''---and what made, and continues to make, it so ``compellingly attractive''.  

\section*{Acknowledgements}

\noindent We thank Prof. Yemima Ben-Menahem (Hebrew University of Jerusalem, IL) and Ms. Abigail Holmes (University of Notre Dame, Indiana, US) for helpful comments and discussion. W.W.~is grateful for support from St.~Cross College, Oxford. P.D.~is grateful for support through the Martin Buber Society of Fellows for Research in the Humanities and Social Sciences, Hebrew University of Jerusalem, IL.

\printbibliography

\end{document}